\documentclass[format=sigconf,screen=true]{acmart} 

\usepackage{setspace}
\usepackage{xspace}
\usepackage{caption}
\usepackage{subcaption}
\usepackage{url}
\usepackage[utf8x]{inputenc}
\usepackage{graphicx}
\usepackage{amsmath}
\usepackage{mathtools}
\usepackage{xcolor,colortbl}

\usepackage{grffile}

\makeatletter
  \newcommand\EatSpacesHack{\@bsphack\@esphack}
\makeatother

\iffalse
  \renewcommand\comment[1]{{\sffamily \color{red} [COMMENT: #1]}}\EatSpacesHack
  \newcommand\xun[1]{{\sffamily \color{purple} \textbf{[xun: #1]}}}\EatSpacesHack
  \newcommand\reviewfix[1]{{\sffamily\color{blue}[FIX: #1]}}\EatSpacesHack
  \newcommand\PostSubmission[1]{}\EatSpacesHack
\else
  \renewcommand\comment[1]{}\EatSpacesHack
  \newcommand\reviewfix[1]{}\EatSpacesHack
  \newcommand\xun[1]{}\EatSpacesHack
  \newcommand\PostSubmission[1]{}\EatSpacesHack
\fi

\newif\ifisanon
\isanonfalse

\newcommand\V[1]{{\mbox{\textit{#1}}}}

\def\Snospace~{\S{}}
\newcommand{\RelaxFloats}{
        \renewcommand{\topfraction}{0.9}
        \renewcommand{\floatpagefraction}{0.9}
        \renewcommand{\textfraction}{0.1}
}
\RelaxFloats

\makeatletter
\newcommand*\ssmall{%
   \@setfontsize\ssmall{7.5}{9.0}%
}
\makeatother

\makeatletter
\newcommand*\sssmall{%
   \@setfontsize\sssmall{5.5}{7.0}%
}

\newcommand{\E}{\mathbb E}
\newcommand{\NA}{--}

\renewcommand\footnotetextcopyrightpermission[1]{} 
\begin{document}

\date{}

%
%

	\title{Peek Inside the Closed World: Evaluating~Autoencoder-Based~Detection of DDoS to Cloud}


	\author{Hang Guo}
	\email{hangguo@isi.edu}
        \affiliation{USC/ISI}

        \author{Xun Fan}
        \email{xufan@microsoft.com}
        \affiliation{Microsoft}

        \author{Anh Cao}
        \email{anhcao@microsoft.com}
        \affiliation{Microsoft}

        \author{Geoff Outhred}
        \email{geoffo@microsoft.com}
        \affiliation{Microsoft}

        \author{John Heidemann}
        \email{johnh@isi.edu}
        \affiliation{USC/ISI}


\begin{abstract}
Machine-learning-based anomaly detection (ML-based AD) has
 been successful at detecting DDoS events in the lab.
However published evaluations of ML-based AD have used only limited data 
 and provided minimal insight into \emph{why} it works.
To address limited evaluation against real-world data,
 we apply autoencoder, an existing ML-AD model,
 to 57 DDoS attack events captured at 5 cloud IPs from a major cloud provider.
We show that our models detect nearly all malicious flows for 2 of the 4 cloud IPs under attack 
 (at least 99.99\%) 
 and detect most malicious flows (94.75\% and 91.37\%) for the remaining 2 IPs.
Our models also maintain near-zero false positives on benign flows to all 5 IPs.
Our primary contribution is 
  to improve our understanding for why ML-based AD works
  on some malicious flows but not others.
We interpret our detection results with feature attribution and counterfactual explanation.
We show that our models are better at detecting malicious flows with anomalies 
 on allow-listed features (those with only a few benign values) than flows with anomalies on deny-listed features (those with
 mostly benign values) because our models are more likely to learn correct normality for allow-listed features.
We then show that our models are better at detecting malicious flows with anomalies on unordered
 features (that have no ordering among their values) than flows with anomalies 
 on ordered features because even with incomplete normality,
 our models could still detect anomalies on unordered feature with high
 recall.
Lastly, we summarize the implications of what we learn
 on applying autoencoder-based AD in production:
 training with noisy real-world data is possible,
 autoencoder can reliably detect real-world anomalies on well-represented unordered features
 and combinations of autoencoder-based AD and heuristic-based filters can help both.
\end{abstract}

\maketitle

\section{Introduction}
	\label{sec:intro}

Anomaly detection (AD) 
 is a popular strategy in detecting DDoS attacks, 
 enabling responses such as filtering.
AD identifies malicious network traffic 
 by profiling benign traffic and flagging traffic deviating 
 from these benign profiles as malicious. 
AD thus assume
  one can profile all benign traffic patterns
  and infer the rest as malicious (the closed world assumption \cite{data_mining}).
Comparing to binary classification,
  another popular strategy in DDoS detection that profiles both benign 
  and malicious traffic and looks for traffic similar to 
  these known malicious profiles,
AD could identify both known and potentially unknown 
 malicious traffic.

Machine learning (ML) techniques
 lead to a new class of DDoS detection study using
 ML-AD models such as one-class SVM~\cite{ocsvm1,ocsvm2,ocsvm3}
 and neural networks~\cite{N-BaIoT,DIOT,Jun01a}.
However, these studies usually suffer from two major weaknesses, 
 limiting their adoption in real-world, operational networks~\cite{Sommer10a}.
First, ML-AD models are often evaluated with limited 
 datasets, often only simulated traffic, traffic from universities or laboratories,
 or two public DDoS datasets (described next).
Prior work has suggested that 
 conclusions based on traffic from simulation and small environments 
 do not generalize to real-world environments at larger scales~\cite{Sommer10a}.
The public datasets from DARPA/MIT~\cite{darpa_dataset} and KDD CUP~\cite{kdd_dataset}
  are synthetic, 20 years old,
  and  have known problems, making them inadequate for contemporary research~\cite{Gates06a,Sommer10a}.
It is thus unclear how well these ML-AD models could detect \emph{real-world} DDoS attacks
 in operational networks. 
Second, prior studies of ML-AD usually do not interpret their 
 models' detection and explain \emph{why} their models work or not work.
Without interpretation, it is difficult for network operators to understand 
 and act on the detection results of ML-based AD systems~\cite{Sommer10a}.
Without explanations on why detection works,
  it is hard to understand the capabilities and limitations
  of ML-based AD in DDoS detection and how one could make the best use of ML-based AD
  in production environment.

Our paper takes steps to addressing these two limitations
  by evaluating ML-based AD with real-world data
  and interpreting the results.

Our first contribution is to evaluate the detection accuracy of autoencoder, 
 an existing ML-AD model, with real-world DDoS traffic from a large commercial
 cloud platform (\autoref{sec:md_cloud_ov}).
We apply our models to 57 DDoS attack events captured from 5 cloud IPs 
 of this platform between late-May and early-July 2019 (\autoref{sec:md_data_ov}).
Detection results show that our models detect almost all malicious attack flows to
 2 of these 4 cloud IPs under attacks (at least 99.99\%) 
 and detect most malicious flows
 for the remaining 2 IPs (94.75\% and 91.37\%, \autoref{sec:rlt_test_accu}).
We show that our models maintain near-zero false positives on benign traffic flows to
 all 5 IPs (\autoref{sec:rlt_det_test_FP}).

Our second contribution is to interpret our detection results with feature attribution (\autoref{sec:md_feat_attr})
 and counterfactual explanation (\autoref{sec:md_cf_ep}) and
 show why our models work on certain malicious flows but not the rest (\autoref{sec:rlt_det_int}).
We show that our models are better at detecting malicious flows with anomalies 
 on allow-listed features (those with only a few benign values) than flows with anomalies on deny-listed features
 (those with mostly benign values)
 because our models are more likely to learn correct normality for allow-listed features (\autoref{sec:rlt_det_int_learn_norm}).
We then show that our models are better at detecting malicious flows with anomalies on unordered
 features (that have no ordering among their values) than flows with anomalies 
 on ordered features because even with incomplete normality,
 our models could still detect anomalies on unordered feature with high
 recall (\autoref{sec:rlt_det_int_det_anom}).
Lastly, our models detect malicious flows with anomalies on packet payload content
 by combining multiple flow features (\autoref{sec:rlt_det_int_muti_anom}).
(We summarize key takeaways from our interpretation results in \autoref{tab:key_int}.)

Out last contribution is to summarize the implications of what we
 learn on using autoencoder-based AD in production (\autoref{sec:imp}):
 training with noisy real-world data is possible (\autoref{sec:imp_noise}),
 autoencoder can reliably detect real-world anomalies on well-represented unordered features
 (whose benign values appear frequently in training data, \autoref{sec:imp_strength})
 and combinations of autoencoder-based AD and heuristic-based filters can help both (\autoref{sec:imp_combine}).

\section{Datasets and Methodology}

Our main contribution is to evaluate ML-based AD with
  real-world data and interpret the results.
Our data is based on a large commercial cloud platform (\autoref{sec:md_cloud_ov}) 
  with real-world DDoS events for several services (\autoref{sec:md_data_ov}).
We then describe our ML-AD models (\autoref{sec:md_ddos_det})
  and the standard techniques we use to
  interpret them (\autoref{sec:md_det_int}).

\subsection{Cloud Platform Overview}
	\label{sec:md_cloud_ov}

We study a large commercial cloud platform
 that is made up of millions of servers 
 across 140 countries.
We study 3 of the wide range of services this cloud platform hosts.
Each of these cloud services is assigned one or more public virtual IPs (VIP).

This cloud platform has seen increasing DDoS attacks over the past years and
 deploys ``in-house'' DDoS detection and mitigation.

In-house detection begins by detecting \emph{DDoS events}
  based on comparing aggregate inbound traffic to an VIP
  to a DDoS threshold.

In-house mitigation employs filtering and rate limiting.
After a DDoS event has been detected,
  each inbound packet to that VIP is
  checked and possibly dropped based on a series
  of heuristics.
These heuristics are filters designed by domain experts
  to identify and filter known DDoS attacks.
Remaining packets are rate limited,
  with any that pass the rate limiter passed to the VIP\@.

The in-house methods consider a DDoS event to end
  when the inbound traffic rate to this VIP goes under the DDoS threshold
  for 
 15 minutes.
In-house mitigation is only applied when there is an ongoing DDoS event (called \emph{war time})
 and is not otherwise applied (during \emph{peace time}).

\begin{figure*}
\centering
\begin{minipage}{.5\textwidth}
  \centering
    \includegraphics[width=0.99\textwidth]{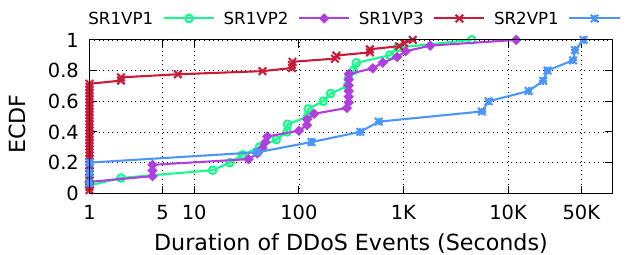}
    \caption{DDoS Events' Durations in Traffic Pcaps}\label{fig:war-dur}
\end{minipage}%
\hfill
\begin{minipage}{.39\textwidth}
  \centering
   \includegraphics[width=0.99\textwidth]{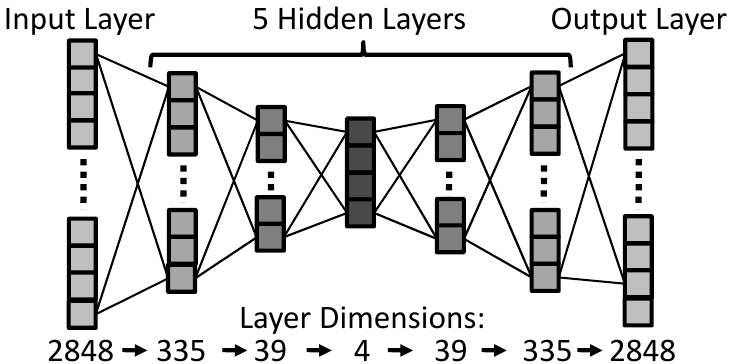}
    \caption{Architecture for Our Autoencoders}\label{fig:autoencoder}
\end{minipage}%
\end{figure*}

\begin{table*}
\setlength\tabcolsep{5pt}
\begin{center}
\small{
\begin{tabular}{c|ccc|rrc|c|c|rr|rr}
&\multicolumn{3}{c|}{\textbf{Total Traffic Pcaps}} & \multicolumn{3}{c|}{\textbf{Total Traffic Flows}} & \multicolumn{1}{c|}{\textbf{Training}} 
& \multicolumn{1}{c|}{\textbf{Threshold}} & \multicolumn{2}{c|}{\textbf{Validation}} & \multicolumn{2}{c}{\textbf{Test}}\\
\textbf{VIPs} & Peace Hrs & War Hrs & DDoS Evts & Benign     & Malicious & DDoS Evts & Benign & Benign & Benign & Malicious & Benign & Malicious \\
 \hline                                                                                      
 SR1VP1 & 110.32 &2.31     &20         & 9,930k     & 119k      & 20 & 1,000k   &   59.5k  &   59.5k  &   59.5k     & 59.5k     & 59.5k    \\  
 SR1VP2 & 96.96  &5.44     &27         & 13,107k    & 1,046k    & 20 & 1,000k   &   523k &   523k &   523k    & 523k    & 523k   \\     
 SR1VP3 & 118.88 &1.36     &49         & 10,704k    & 90k       & 7  & 1,000k   &   45k  &   45k  &   45k     & 45k     & 45k     \\
 SR2VP1 & 57.73  &58.89    &15         & 5,469k     & 37k       & 10 & 1,000k   &   18.5k  &   18.5k  &   18.5k     & 18.5k     & 18.5k     \\  
 SR3VP1 & 182.99 &0        &0          & 698k       & 0         & 0 &  548k    &   50k  &   50k  &   0       & 50k     & 0       \\ 
\end{tabular}
}
\end{center}
\caption{Summary of Traffic Pcaps and Traffic Flows Used in This Paper}
        \label{tab:data_sum}
\end{table*}

\subsection{Cloud DDoS Data}
	\label{sec:md_data_ov}

To evaluate 
  ML-based AD, 
  we obtain peace and war-time traffic packet captures (\emph{pcaps})
  from VIPs on this cloud platform
  and extract benign user traffic and malicious DDoS traffic.

\textbf{Cloud VIPs:}
We study five VIPs from  three different services hosted on this cloud platform (see \autoref{tab:data_sum} for anonymized VIPs).
Three of these VIPs (SR1VP1, SR1VP2 and SR1VP3) are instances of a gaming communication service (SR1), 
  each in a different data center and physical location.
The other two VIPs (SR2VP1 and SR3VP1) belong to two different gaming authentication
  services (SR2 and SR3).

\textbf{Traffic Pcaps:}
We obtain over 100 hours of inbound traffic pcaps to each of these 5 VIPs.
Each VIP's pcaps include all war-time traffic and partial peace-time traffic,
  observed at this VIP in a 8-day period between late-May and early-July 2019
(specific times in \autoref{tab:data_sum}).
We use partial peace-time traffic because we find more traffic does not
 increase our models' accuracy.
We observe SR3VP1 for extended 180 hours because this VIP 
 receives less traffic than other VIPs.
Our pcaps are sampled, retaining 1 in every 1000 packets.

Our 5 VIPs see different distributions of DDoS events in this period, as in \autoref{fig:war-dur}.
SR1VP3 sees a large number (49) of mostly short DDoS events (71\% being
 1 second or less, see red crosses in \autoref{fig:war-dur}).
The cloud platform's DDoS team suggests these brief DDoS events 
  are likely 
  botnets randomly probing IPs.
In comparison, SR1VP1 and SR1VP2 see smaller numbers of 
 longer DDoS events,
  with median durations of 121 and 
  140 seconds for their 20 and 27 events (\autoref{fig:war-dur}).
SR2VP1 is frequently attacked,
  with about 59 hours of war time,
  and sees DDoS events of broad range of durations (from 1 second to more than 14 hours).
The cloud's DDoS team reports that this VIP is hosting a critical 
 service, 
 so long attacks are likely attempts to gain media attention.
SR3VP1 reports no DDoS events since 
 service SR3 is rarely attacked.
We use SR3VP1 to evaluate false positives with our detection methods.

\textbf{Benign and Malicious Traffic:}
We report peace-time traffic as
  ``benign traffic''.
While there may be very small attacks in the peace-time traffic,
  the cloud platform considers any such events too small to
  impact the service
  and does not filter them.
(We choose to not remove such attacks to evaluate our system on noisy, real-world traffic~\cite{Gates06a}.)
War-time traffic is also a mix of benign user traffic and malicious DDoS traffic.
We only consider the fraction of war-time traffic dropped by heuristic-based
 filters from in-house mitigation as malicious (annotated as ``malicious traffic'' hereafter), 
 recalling these heuristics
 identify known attacks (\autoref{sec:md_cloud_ov}).
We only use these malicious traffic to evaluate our methods and ignore the rest of war-time traffic
  since we do not have perfect ground truth for them.

\textbf{Benign and Malicious Flows:}
 %
Since our models' detection relies on 
  flow-level statistics 
  like packet counts and rates,
we first aggregate packets from benign and malicious traffic as 5-tuple flows.

We summarize the number of benign and malicious flows and number of DDoS
 events in these malicious flows in \autoref{tab:data_sum}.
Since our malicious flows come from a subset of war-time traffic
 that matches in-house mitigation's heuristics,
 the DDoS events in malicious flows are a subset of
 DDoS events in war-time traffic pcaps.

Our data is predominantly UDP (99.87\% of our 40M flows in \autoref{tab:data_sum}),
 likely due to all three cloud services we study
 are latency-sensitive gaming services.
We have not evaluated if our results apply to TCP-based services.
Future work may relax this limitation.

\textbf{Extracting Flow Features:}
We use Argus~\cite{argus}
 to extract 23 flow features (see \autoref{tab:data_flow_feat})
 from the first 10 seconds (an empirical threshold) of 
 each benign and malicious flow.

Our 23 flow features can be categorized into two groups.
The first group of features (ports, rates, and packet sizes, gray in \autoref{tab:data_flow_feat})
  are those used in in-house mitigation's heuristics.
These features enables us to understand if our models
 use the same features in detecting certain malicious flows
 as in-house mitigation does. 
Other features
 (such as flow inter-packet arrival time and packet TTLs, white columns in \autoref{tab:data_flow_feat})
 are 
 not used by in-house mitigation, likely because they are
 less intuitive to humans
These features enables us to explore
 how well ML-AD models could compliment human expertise
 by using more subtle features in detection.

While many of our features, such as 
 packet rate (\autoref{tab:data_flow_feat}),
 are distorted by data sampling, we believe our detection still works 
 because our models are trained to identify sampled traffic.

\textbf{Unordered Feature Encoding:}
Since three of our features (source port, destination port and protocol) are unordered 
 and directly using them 
  would implicitly create an ordering among their values
  (for example, implying that port 5 is more similar to port 6 than port 4 is),
we use one-hot encoding~\cite{one-hot-intro} to avoid this distortion.
We map protocol into 256 one-hot features (is\_protocol\_0, is\_protocol\_1, ... is\_protocol\_255),
  each with a binary value.
Similarly, we map ports into 1286 one-hot features, each representing a group of 51 adjacent ports (1 to 51, 52 to 102, ... 65485 to 65535), 
  with port 0 used to indicate both
  illegal TCP/UDP port 0 and non-existent port in non-TCP-UDP flows.
(We group every 51 ports because otherwise we will need 65536$\times$2 one-hot 
  features to represent 
 source and destination ports, more than our machine can handle.)
Grouping ports could cause false positives or negatives if two common ports
  appear in the same aggregate, we examined our data and found that all popular ports differ by at 
  least 53 in the port space and we never group popular ports.

\begin{table*}
\setlength\tabcolsep{5pt}
\begin{center}
\ssmall{
\begin{tabular}{l|l|l|l|l|l|l|l|l|l|l}
\hline
\textbf{\cellcolor{gray!12}Sport} & \textbf{\cellcolor{gray!12}Dport} & \textbf{\cellcolor{gray!12}Proto}  & \textbf{SrcPkts} & \textbf{\cellcolor{gray!12}SrcRate} & \textbf{SrcLoad}  & \textbf{SIntPkt}        & \textbf{sTtl}  & \textbf{\cellcolor{gray!12}sMaxPktSz} & \textbf{\cellcolor{gray!12}sMinPktSz} &  \textbf{SrcTCPBase}\\
\hline
\cellcolor{gray!12}source         & \cellcolor{gray!12}dest           & \cellcolor{gray!12}protocol        & src-to-dst       &\cellcolor{gray!12} src-to-dst       & src-to-dst       &  mean src-to-dst inter   & TTL in last src &  \cellcolor{gray!12}src-to-dst        & \cellcolor{gray!12}src-to-dst          &   src TCP base    \\
\cellcolor{gray!12}port           & \cellcolor{gray!12}port           & \cellcolor{gray!12}number          & pkt count        &\cellcolor{gray!12} pkt/s            & bits/s       &  -pkt arrival time       & -to dst pkt     & \cellcolor{gray!12} max pkt size      & \cellcolor{gray!12}min pkt size        &   sequence          \\  
\hline 
\multicolumn{11}{c}{\textbf{TcpOpt\_\{M,~w,~s,~S,~e,~E,~T,~c,~N,~O,~SS,~D\}}}  \\
\hline 
\multicolumn{11}{l}{the existence of certain TCP option: max segment size (M), window scale (w), selective ACK OK (s), selective ACK (S), TCP echo (e), TCP echo reply (E),}\\
\multicolumn{11}{l}{TCP timestamp (T), TCP CC (c), TCP CC New (N), TCP CC Echo (O), TCP src congestion notification (SS) and TCP dest congestion notification (D)} \\
\hline
\end{tabular}
}
\end{center}
\caption{Our 23 Flow Features (Merging 12 Features About Existence of Certain TCP Option) Before One-hot Encoding}
        \label{tab:data_flow_feat}
\end{table*}

\subsection{DDoS Detection Techniques}
	\label{sec:md_ddos_det}

Having obtained benign and malicious flows,
 we next describe the ML models we use 
 and how we train, validate and test these models with these flows.
We developed our specific ML-based AD techniques ourselves,
 but we follow the use of autoencoder like prior work~\cite{N-BaIoT,autoencoder_anomdet1,autoencoder_intrution1,autoencoder_outlier1}
 and we specifically follow the idea of N-BaIoT of using reconstruction error to detect DDoS~\cite{N-BaIoT}.
Our goal is not to show a new detection method, but to evaluate 
 and interpret current state-of-the-art methods with real world data.

\textbf{Model Overview:}
We use a type of neural-network ML model called autoencoder
 because it is widely used in AD
 (such as 
  system monitoring~\cite{autoencoder_anomdet1} 
 and outlier detection~\cite{autoencoder_outlier1}) 
 and has been shown to detect DDoS attacks accurately 
 in lab environment (\cite{N-BaIoT}).
While other ML models are also used for AD, 
 such as one-class SVM~\cite{ocsvm1,ocsvm2,ocsvm3}
 and other neural networks~\cite{Chalapathy18a,Oza19a,DIOT,Jun01a}.
We currently focus on autoencoder and leave studying other models
 for future work.

Autoencoder is a symmetric neural network that
 reconstructs its input by compressing the input to a smaller dimension
 and then expanding it back~\cite{autoencoder-intro}.
The aim of autoencoder is to minimize
  reconstruction error,
  the differences
   between input and output (the reconstructed input).
We compute the difference between input and output vectors ($F_{in}$ and $F_{out}$)
 as the mean of element-wise square error, as shown in \autoref{eqn:mse}
 where $N$ is the number of elements in $F_{in}$ and $F_{out}$; 
 and $F_{in}^i$ (or $F_{out}^i$) is the i-th element in
 $F_{in}$ (or $F_{out}$). 

\begin{align}\label{eqn:mse}
        E(F_{in},F_{out})=\frac{\sum_{i=1}^{N}(F_{in}^i-F_{out}^i)^2}{N}
\end{align}


To detect malicious DDoS flows, 
 we train an autoencoder with only benign flows
 and identify malicious flows by looking for
 large reconstruction errors.
The rationale is
  the autoencoder
  learns
  to recognize useful patterns in the benign flows
  with, in-effect, lossy compression.
When it encounters statistically different flows like the malicious flows, 
  it cannot compress this anomalous traffic efficiently 
  and so
  produces a relatively large reconstruction error,
  with the degree of error reflecting the deviation from normal of the anomaly.

We build a 6-layer neural network 
 for each of our 5 VIPs,
 compressing a 2848-by-1 input vector (2$\times$1286 one-hot features for ports, 256 one-hot features for protocols
  and the other 20 features in \autoref{tab:data_flow_feat}) to a 4-by-1 vector
 and expand it back symmetrically (dimensions of each layer shown in \autoref{fig:autoencoder}).
As with many ML systems, the specific choices of 4-by-1 and 6 layers
  are empirical, although we also tried 8 layers without seeing much advantage.
We use ReLu~\cite{relu} as activation function, 
  L2 regulation~\cite{pytorch-weight-decay} and
  dropout~\cite{dropout} to prevent overfitting
and mini-batch Adam gradient descent~\cite{adam} for model optimization, 
  all following standard best practices~\cite{nn_course}.
Our implementation uses 
 pyTorch~\cite{pyTorch}.

\textbf{Model Training:}
We train each VIP's autoencoder to 
 accurately reconstruct benign flows from this VIP.

We first randomly draw 1 million benign flows from each VIP as its training
 dataset (see ``training'' column of \autoref{tab:data_sum}).
SR3VP1 observes only 698k benign flows, even with extended observation,
  so there we train on 548k benign flows.
(We experimented with additional training data but did not find it helped)

We then pre-process training dataset by normalizing training flows' feature values to approximately
  the same scale (about 0 to 1), following best practices~\cite{nn_course}.
The one-hot features are already normalized,
  but for a given other feature $i$ of flow $w$ in
  the training dataset ($F^i_w$ in \autoref{eqn:min_max_norm}), 
  we normalize it with min-max normalization (\autoref{eqn:min_max_norm} where $F^i_{tmax}$ and $F^i_{tmin}$ are
 the maximum and minimum values for feature $i$ in all training flows).


We initialize four hyper-parameters in our models: mini-batch size as 128, 
  learning rate as $10^{-5}$, drop-out ratio as 50\% (per recommendation~\cite{dropout}) and weight decay for L2 regulation (\cite{pytorch-weight-decay})
  as $10^{-5}$.
(We tune these values during model validation below if needed.)

Lastly, we train our models with normalized training data for 2 epochs.
(Adding more epochs does not increase 
 models' detection accuracy on validation datasets, 
  and risks overfitting.)

\textbf{Threshold Calculation:}
Detecting malicious flows from large reconstruction error
  requires a threshold to separate normal error from anomalies.
We calculate this threshold by estimating the upper bound for benign flows' errors.
We randomly draw benign flows from each VIP 
  to form threshold datasets
  (see ``threshold'' column of \autoref{tab:data_sum}).
We set the size of threshold dataset to match the size of validation and test dataset (described later this section).
Similar to model training, we pre-process threshold data with min-max normalization (\autoref{eqn:min_max_norm})
 and maximum and minimum feature values extracted from training datasets ($F^i_{tmax}$ and $F^i_{tmin}$).
We apply trained models to flows in threshold dataset and record their reconstruction
 errors as $\E$.
We calculate detection threshold 
 with \autoref{eqn:det_thre}
 where $\mu_{\E}$ and $\sigma_{\E}$ are mean and standard deviation of $\E$.



\noindent\begin{minipage}{.24\textwidth}
\begin{align}\label{eqn:min_max_norm}
    \hat{F^i_w}=\frac{F^i_w-F^i_{tmax}}{F^i_{tmin}-F^i_{tmax}}
\end{align}
\end{minipage}
\vspace{0.15em}
\hfill
\noindent\begin{minipage}{.24\textwidth}
\begin{align}\label{eqn:det_thre}
    T_{det}=\mu_{\E}+3\sigma_{\E}
\end{align}
\end{minipage}
\vspace{0.15em}

\textbf{Model Validation:}
We validate detection accuracy of trained models (with initial hyper-parameters) by applying them
 to detect benign and malicious flows in validation datasets.
When we encounter poor accuracy in the validation data,
   we tune hyper-parameters of the models 
  to improve validation accuracy.

To validate our model,
  we construct validation dataset for each VIP by randomly drawing half malicious flows
 from a VIP and equal amount of random benign flows from same VIP (shown under ``validation'' of \autoref{tab:data_sum}).
We pre-process validation dataset with min-max normalization and $F^i_{tmax}$ and $F^i_{tmin}$ (\autoref{eqn:min_max_norm}).
We apply trained models to detect benign and malicious flows in validation sets and check common accuracy metrics of detection results:
 mainly precision ($\V{TP}/(\V{TP}+\V{FP})$), recall ($\V{TP}/(\V{TP}+\V{FN})$) and F1 score (${2\times\V{precision}\times\V{recall}}/({\V{precision}+\V{recall}})$)
  where $\V{TP}$, $\V{FP}$ and $\V{FN}$ stands for true positives, false positives and false negatives in identifying malicious flows.
Note that for SR3VP1 where we only have benign flows, we instead examine its true negative ratio (TNR, the fraction
 of benign flows that get correctly detected.)

If any detection metric for a per-VIP model goes under 99\%,  
 we tune this model's hyper-parameters
  with random search~\cite{random_search},
  by training multiple versions of this model, each with a set of randomly-chosen values for hyper-parameters.
We then select as the final model the version that gets the highest F1 score 
  against the validation dataset and use this final model for all subsequent detection.
(\autoref{tab:final_para_val} lists hyperparamter values for our final models.)


\textbf{Model Testing:}
Finally, we report detection accuracy for our trained and validated models
 by applying them to test datasets, consisting of
 the other half of malicious flows extracted from each VIP
 and equal amount of random benign flows from the same VIP (see ``Test'' of  \autoref{tab:data_sum}).
Specifically, we first pre-process test dataset with min-max normalization and $F^i_{tmax}$ and $F^i_{tmin}$ (\autoref{eqn:min_max_norm}).
We then report our models' detection  precision, recall and F1 score on test dataset.
(Similar to validation, we report TNR for SR3VP1.)

\subsection{Techniques to Interpret Detections}
	\label{sec:md_det_int}

While our models follow best practices,
  we are the first to evaluate such models with real-world data
  and interpret the results.
We interpret our models' detection results with feature attribution (\autoref{sec:md_feat_attr}) and 
 counterfactual explanations analysis (\autoref{sec:md_cf_ep}).

\subsubsection{Feature Attribution}
	\label{sec:md_feat_attr}
We use feature attribution analysis to understand the contribution
 from each feature to the detection of each flow instance.
Prior work used feature attribution~\cite{Zeiler13a,zintgraf17a,simonyan13a,shrikumar16a}.
They either attribute feature importance by 
  evaluating the difference in model output when perturbing each input feature (\cite{Zeiler13a,zintgraf17a}),
  or by taking the partial derivative of model output to each input feature (\cite{simonyan13a,shrikumar16a}).

\begin{align}\label{eqn:feat_attr}
        A(j)=\frac{(F_{in}^j-F_{out}^j)^2}{\sum_{i=1}^{N}(F_{in}^i-F_{out}^i)^2}
\end{align}

Since our models' detection is based on reconstruction
 error of input flow (\autoref{eqn:mse}), 
  which is the mean of per-feature errors from all flow features, 
  we can measure a feature's contribution
  to detection by how much error it contributions to
  overall reconstruction error.
We normalize per-feature error
 by dividing it with the sum of error from all features, as in \autoref{eqn:feat_attr},
  and attribute that feature's contribution
  as this normalized per-feature error.

\subsubsection{Counterfactual Explanations}
	\label{sec:md_cf_ep}

Counterfactual explanations show how an input must
 change to significantly change its detection output,
 as advocated by prior work~\cite{wachter17a,Martens14a}.
We use counterfactual explanations to understand the normality our models
  learn for each flow feature, 
  suggesting values the models consider anomalous.

Specifically, we first find a \emph{base flow} that is detected as benign,
  then we repeatedly alter the target feature's value in this
  base flow while keeping other features unchanged.
We feed these altered base flows into our model
  to observe how much the reconstruction error changes
  with each perturbation of target feature's value: an increase
 in errors suggests our models consider current feature value more abnormal than the previous value,
 and vice versa.
%
We repeat this experiment on different base flows to see if
our models consistently consider certain target feature values more normal
 than the other values,
  with relatively normal values suggesting normality our models learned.

\section{Detection Results}
	\label{sec:rlt}
\begin{table}
\setlength\tabcolsep{2.5pt}
\small{
\begin{tabular}{r|cccc}
                         & Mini-batch Size &  Learning Rate & Drop-out Ratio & Weight Decay \\
 \hline
 SR1VP1                               &  64             &   $2\times10^{-5}$ & 10\%    & $10^{-6}$       \\
 SR2VP1                               &  32             &   $10^{-5}$        & 10\%    & $2\times10^{-6}$ \\
 Other VIPs                      &  128            &   $10^{-5}$        & 50\%    & $10^{-5}$       \\
\end{tabular}
}
\caption{Hyperparameters Values for Final Models}
        \label{tab:final_para_val}
\end{table}

%

To understand how well ML-based AD works in detecting real-world DDoS attacks,
 we train and validate an autoencoder model for each of our 5 VIPs 
  as described in \autoref{sec:md_ddos_det}.
We summarize hyperparameters values for our final models in \autoref{tab:final_para_val}
 where 
 models for SR1VP1 and SR2VP1 use tuned hyperparameters values 
 and models for other 3 VIPs use initial hyperparamter values
  from \autoref{sec:md_ddos_det}.

With trained and validated models,
  we report detection accuracy on test datasets in \autoref{sec:rlt_test_accu} and 
  examine false positives in \autoref{sec:rlt_det_test_FP}.
In \autoref{sec:rlt_det_by_anorm},
  we evaluate our models on all malicious flows
  (recalling test datasets only contain half of total malicious flows)
  and interpret \emph{why} our models detect some malicious flows
  but miss others in \autoref{sec:rlt_det_int}.


\subsection{Detection Accuracy on Test Dataset}
	\label{sec:rlt_test_accu}

We evaluate accuracy by measuring
  precision, recall, F1 score and TNR of our models' detection
  of test datasets in \autoref{tab:overall_det_accur}.

We first observe that our model's detection precision 
 and TNR for all 5 VIPs are high (at least 98.90\% in \autoref{tab:overall_det_accur}),
 suggesting they rarely generate false alerts: only 2,556 (0.36\%) false positives out of
 all 696,000 tests of benign flows.
(We later show that only 28 of these 2,556 are actual false positives in \autoref{sec:rlt_det_test_FP}.)

\begin{table}
\begin{center}
\small{
\begin{tabular}{r|ccccc}
           & SR1VP1  & SR1VP2 & SR1VP3 & SR2VP1 & SR3VP1      \\
 \hline
 Precision &  98.90\%        & 99.69\%          & 99.81\%           & 99.50\%       & \NA  \\
 Recall    &  94.75\%        & 99.99\%          & 100.0\%           & 91.37\%       & \NA \\
 F1-Score  &  96.78\%        & 99.83\%          & 99.90\%           & 95.26\%       & \NA \\
 TNR       &   \NA           & \NA              &\NA                & \NA           &  99.68\% \\
\end{tabular}
}
\end{center}
\caption{Detection Accuracy on Test Dataset}
        \label{tab:overall_det_accur}
\end{table}

Our second observation is that our models identify almost all malicious flows 
 to 2 of the 4 VIPs under attack (detection recall is 99.99\% for SR1VP2 and 100\% for SR1VP3)
 and identify most malicious flows to the other 2 VIPs (recall is 94.75\% for SR1VP1 and 91.37\% for SR2VP1),
 as shown in \autoref{tab:overall_det_accur}.


\subsection{Examining False Positives on Test Dataset}
        \label{sec:rlt_det_test_FP}

Our models make 2,556 false positives against the test datasets (\autoref{sec:rlt_test_accu});
  we next compare these to in-house mitigation's heuristics
  such as allow-lists of destination ports and protocols.

We first show most of these false positives (95.7\%, 2,446 out of 2,556, see \autoref{tab:FP-breakdown})
 are actually true positives (correctly-detected malicious flows),
 recalling our noisy training data may contain some malicious traffic (\autoref{sec:md_data_ov}).
We find 
 most of these actual true-positive flows
 (79.8\%, 1,953 out of 2,446) are UDP flows with malicious destination ports.
We also find a small fraction of them 
  using malicious source ports (0.2\% or 4),
  and a few with at least one packet with bad payload content
  (that fails regular expressions required by in-house mitigation's heuristics) (0.1\% or 2).
(We show in \autoref{sec:rlt_det_int_muti_anom} that our models could detect some malicious flows with bad packet payload content
  based on anomalies in flow features.)

We next show a few false positives (3.2\%, 82 out of 2,446) are artifacts 
  due to misdirected TCP flows (\autoref{tab:FP-breakdown}).
These misdirected flows appear to originate from our 5 VIPs,
  yet the pcaps we study contain only inbound packets to these VIPs (\autoref{sec:md_data_ov}).
These misdirected flows thus have wrong values of zeros 
 for some of our features such as source-to-destination packet counts
 and rates (\autoref{tab:data_flow_feat}).
We confirm that these flows' directions
 are actually mis-labeled 
 due to a known limitation of Argus.

Lastly, we show the remaining 28 false positives
  are likely actual false positives.
Each of them (all TCP flows) does not match any of in-house mitigation's heuristics.

We conclude that only a tiny fraction of false positives reported in \autoref{sec:rlt_test_accu},
 are actual false positives (1.1\%, 28 out of 2,556),
 suggesting the actual false positive rate is near zero (0.00\%, 28 of 696,000 test benign flows).
(We explore the potential causes for these actual false positives in \autoref{sec:rlt_det_int_det_anom}.)

\begin{table}
\small{
  \begin{tabular}{lrrr}
    total false positives                                   & 2,556    & (100.0\%)&  \\
    \qquad  actual false positives                          & 28       & (1.1\%)  & \\
    \qquad  actual true positives                           & 2,446    & (95.7\%) & (100.0\%) \\
    \qquad\qquad UDP flows w bad dst port       & 1,953    & (76.5\%) & (79.8\%)\\
    \qquad\qquad UDP flows w bad src port           & 4        & (0.2\%) & (0.2\%)\\
    \qquad\qquad UDP flows w bad payload content            & 2        & (0.1\%) & (0.1\%)\\
    \qquad\qquad flows w bad protocols          & 487      & (19.1\%) & (19.9\%) \\
    \qquad misdirected TCP flows                            & 82       & (3.2\%)  & \\
  \end{tabular} 
}
  \caption{False Positives on Test Dataset Breakdown}
                \label{tab:FP-breakdown}
\end{table}

\subsection{Detection Accuracy On All Malicious Flows}
        \label{sec:rlt_det_by_anorm}

We next explore how well our models detect all malicious flows we have,
 recalling test datasets contain only half of them (\autoref{sec:md_ddos_det}).
We group malicious flows by their main anomalies
  as detected in-house mitigation, 
  and show which anomalies are best detected by our models,
  and which are poorly detected.

Our models are near perfect at detecting anomalies 
 on allow-listed features (those with mostly malicious values besides
 a few benign values, judged by in-house mitigation’s heuristics) 
 with unordered values: destination port and protocol.
As a result, our models capture all flows with malicious protocol (100.00\% of about 15k) 
 and nearly all UDP flows with malicious destination ports (99.92\% of about 1M),
 see \autoref{tab:recall_by_anom}. 

Our models are reasonable at detecting anomalies on deny-listed features
 (those with mostly benign values besides a few malicious values, judged by in-house mitigation’s heuristics) 
 with unordered values: source port.
Our models identify most UDP flows with malicious source ports (97.5\% of 5k). 

However, we find our models are bad at detecting anomalies on deny-listed features 
 with ordered values: flow packet sizes.
Our models detect only a few malicious flows (8.5\% of about 3k)  
 containing packets with too-small payload (in-house mitigation drops
 UDP packets with payload smaller than a threshold), as in \autoref{tab:recall_by_anom}.
(In \autoref{sec:rlt_det_int_learn_norm}, we show 
 that our models 
 infer if a UDP flow contains packets with too-small payloads
 based on feature  
 maximum and minimum flow packet size, recalling \autoref{tab:data_flow_feat}.)

Lastly, our models detect a quarter of UDP flows (24.7\% of 8k) 
 containing 
 packets with bad payload contents (that fail
 regular expressions required by in-house mitigation), despite our models do not see
 packet payloads (\autoref{tab:recall_by_anom}).

\begin{table}
\setlength\tabcolsep{3.5pt}
\begin{center}
\small{
\begin{tabular}{lr|rr}
\multicolumn{2}{c|}{\textbf{Total Flows by Main Anomalies}} & \multicolumn{2}{c}{\textbf{Detected Flows }} \\
Main Anomaly                         & Count                   & Count              &  Frac of Total          \\
 \hline
Flows w Bad Protocol      & 15,206                  &15,206              &100.00\%           \\
UDP Flows w Bad Dst Port  & 1,261,951               &1,260,943           &99.92\%            \\
UDP Flows w Bad Src Port      & 5,334                   &5,201               &97.5\%              \\
UDP Flows w Too Small Payload         &2,522                    &215                 &8.5\%       \\
UDP Flows w Bad Payload Contents      &8,229                    &2,036               &24.7\%        \\
\end{tabular}
}
\end{center}
\caption{Detection to All Malicious Flows Breakdown}
        \label{tab:recall_by_anom}
\end{table}


\section{Interpreting Detection of Malicious Flows}
	\label{sec:rlt_det_int}

\begin{table*}
\small{
 \begin{tabular}{c|l}
\hline
 \textbf{No.} & \textbf{Descriptions} \\
\hline
  1 & Autoencoder can be trained successful with noisy data (\autoref{sec:rlt}), provided all
     benign values of target feature appear frequently in this data (\autoref{sec:rlt_det_int_learn_norm}) \\
 \hline
  2 & Autoencoder can reliably detect anomalies on features whose benign values are 
     frequent among training data (\autoref{sec:rlt_det_int_learn_norm}) and who are also unordered (\autoref{sec:rlt_det_int_det_anom}) \\
 \hline
  3 & Autoencoder almost always combines anomalies from multiple features in detection (\autoref{sec:rlt_det_int_muti_anom}), using even features that are less intuitive for human (\autoref{sec:rlt_det_int_det_anom}) \\
 \hline
  \end{tabular}
}
  \caption{Key Takeaways from Interpretation Results}
                \label{tab:key_int}
\end{table*}


A contribution of our work is to interpret
  why ML-based AD detects certain anomalies better than others.
We show that our models are better at detecting anomalies
 on allow-listed features than those on deny-listed features
 because they are more likely to learn correct normality for 
 allow-listed features (\autoref{sec:rlt_det_int_learn_norm}).
Our models are better at detecting anomalies on unordered
 features than those on ordered features because even with incomplete normality,
 our models could still detect anomalies on unordered feature with high 
 recall (\autoref{sec:rlt_det_int_det_anom}).
Lastly,
 our models detect anomalies on packet payload content 
 by combining multiple flow features (\autoref{sec:rlt_det_int_muti_anom}).

We summarize key takeaways from our interpretation results in \autoref{tab:key_int} and
  describe our interpretation results' implications on using autoencoder-based AD in production later in \autoref{sec:imp}.

\subsection{Learning Normalities for Features}
	\label{sec:rlt_det_int_learn_norm}
We show our models are more likely to learn correct normality
 for allow-listed features (destination port and protocol)
 than for deny-listed features (source port and flow packet sizes).
The rationale is that
 since our models learn frequently seen values in training data as normality,
 it is more likely for allow-listed features to have all their benign values 
 frequently seen in training data and thus learned as normality
 because they have, by definition (\autoref{sec:rlt_det_by_anorm}), 
 only a few benign values.
(We show how the normalities learned affect our detection of anomalies later in
 \autoref{sec:rlt_det_int_det_anom}.)



\textbf{Allow-listed Destination Port:}
We show our models correctly learn normality for destination port
(with all values being malicious except one benign value, per in-house mitigation's heuristics)
 because the one benign port is the most frequent among training data.

We explore the normality our models learn for destination ports
 with counterfactual explanation (\autoref{sec:md_cf_ep}).
We randomly draw 100 UDP flows, detected as benign,
  from each VIP's test datasets as base flows, alter these 500 base flows by enumerating their
  destination ports from 0 to 65535 with a step size of 51 (0, 51, ... 65535) and feed
  altered flows into models.
The step size is because our models merge each 51 adjacent ports to one feature (\autoref{sec:md_data_ov}).
We then watch for how base flows' errors change as destination ports change.

\begin{figure}
\centering
  \includegraphics[width=0.48\textwidth]{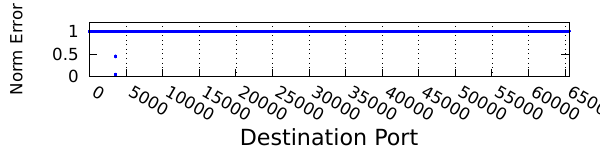}
    \caption{Normalized Reconstruction Errors for 100 Base Flows from SR1VP2 using Different Destination Ports}\label{fig:dist-reon-error-dports1v2}
\end{figure}

We show our models correctly learn the normality of destination ports
 by consistently considering base flows with malicious ports more abnormal than the
 same flows with the one benign port.
We use reconstruction errors from SR2VP1's 100 base flows as example (other VIPs are similar).
Since we only care about how a base flow's error changes as its destination port changes (rather
 than the exact values of these errors)
 and want to compare these changes across all 100 base flows from this VIP,
we normalize the set of errors resulted from one base flow using different destination ports
 to range $[0, 1]$ by dividing these errors with the maximum error found among them.
We plot normalized errors for SR2VP1's 100 base flows with different destination ports
 as blue dots in \autoref{fig:dist-reon-error-dports1v2}.
We show that all malicious ports lead to similarly high
 reconstruction errors and this pattern is very consistent across all 100 base flows from SR2VP1
 (represented by the horizontal blue line at normalized error 1 in \autoref{fig:dist-reon-error-dports1v2}).
We also show consistently low errors (at most 0.46) at the one benign port for SR2VP1
 (shown as the blue dots to the left of port 5000 and below error 0.5 in \autoref{fig:dist-reon-error-dports1v2}).

\textbf{Allow-listed Protocol:}
%
We show our models learn incomplete normality for protocols.
Per in-house mitigation's heuristic,
 all protocols are malicious except UDP (for SR1, SR2 and SR3), TCP (for SR1 and SR3)
 and 3 other protocols (for SR3 only, exact protocols omitted for security).
However our models only learn UDP as benign
 because the other 4 protocols are infrequent in training data.

We explore normality our models
 learn for protocols by
 applying counterfactual explanation to the same 500 base flows from destination port analysis, varying
 their protocols from 0 to 255 (with a step size of 1)
 and watch for how their reconstruction error changes.

\begin{figure}
\centering
  \includegraphics[width=0.48\textwidth]{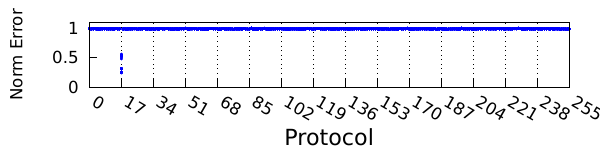}
    \caption{Normalized Reconstruction Errors for 100 Base Flows from SR3VP1 using Different Protocols}\label{fig:dist-reon-error-proto3v1}
\end{figure}

We show our models learn incomplete normality for protocols
  by consistently considering based flows with non-UDP protocols
  more abnormal than same flows with UDP.
For example, in SR3VP1's normalized errors (\autoref{fig:dist-reon-error-proto3v1}), 
 we find blue dots representing UDP (at protocol 17)
 consistently 
 correspond to low errors 
(at most 0.56).
(Other 4 VIPs are similar.)

We believe 
 that our models fail to learn non-UDP benign protocols
 because they are 
 infrequent in training data.
While UDP accounts for almost all training data for our 5 VIPs (99.87\% of 4.5M), 
 TCP accounts for only a tiny fraction (0.01\% of 4.5M) 
 and the 3 other benign protocols for SR3 are completely missing.
We note that TCP flows show up even less than noises in training data
 (flows with malicious protocols, showing up in 0.11\% of 4.5M), 
 suggesting that it actually makes sense for our models to ignore 
 infrequent benign protocols
 like TCP otherwise it risks learning noises.



\begin{figure}
\centering
  \includegraphics[width=0.48\textwidth]{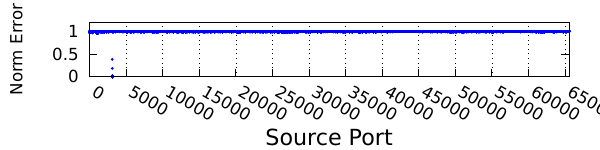}
    \caption{Normalized Reconstruction Errors for 100 Base Flows from SR2VP1 using Different Source Ports}\label{fig:dist-reon-error-sports2v1}
\end{figure}

\textbf{Deny-listed Source Port:}
We show our models are unlikely to learn correct
 normality for deny-listed features like source port because
 their values are mostly benign (per definition \autoref{sec:rlt_det_by_anorm})
 and it is unlikely for all their benign values
 to be frequently seen in training data
 and learned as normality.

We explore normalities our models learn for source port by
 applying similar counterfactual explanation analysis
 as we do for destination port.



We show our models fail to learn the correct normality for source ports.
Per in-house mitigation's heuristic,
 most source ports are benign (besides 1024 and 1023 malicious ports) for SR1 and SR2
 and all source ports are benign for SR3.
However, our models only consider 
 a few source ports frequently seen in training data (``frequent training ports'')
 as relatively normal.
As an example, we show reconstruction errors of SR2VP1's 100 base flows in \autoref{fig:dist-reon-error-sports2v1}.
(We summarize results for other 4 VIPs later.)
We find for SR2VP1, source port 3111 (blue dots left of port 5000 and below
 error 0.5 in \autoref{fig:dist-reon-error-sports2v1}) consistently lead to low errors (at most 0.38)
 while the rest ports 
 lead to high errors (see horizontal blue line at error 1).
We believe the reason for port 3111's low errors
 is that it 
 corresponds to benign source port 3074 which is the most frequent among SR2VP1's training
 data, (in 75.31\% of 998k training UDP flows), 
 considering our models do not distinguish among port 3061 to 3111
 due to our grouping of adjacent 51 ports during one-hot encoding (\autoref{sec:md_data_ov}).
We find similar trend in reconstruction errors for other 4 VIPs: low error with
 the most frequent training source port (all benign) 
 and high errors with the rest source ports.
The only exception is that we find one malicious source port for SR1VP1 (omitted for security)
 also leads to low error  
 due to it is the second most frequent 
 among SR1VP1's training data  
 (in 1.21\% of 999k UDP training flows, 
 recalling our training data is noisy).
We conclude that our models learn incorrect normality for SR1VP1 
 (considering one benign and one malicious source ports normal) 
 and incomplete normality for other 4 VIPs (considering one benign port normal).

\textbf{Deny-listed Packet Sizes:}
Similarly, our models are not likely to learn correct normality
 for deny-listed flow packet sizes.

Our models detect malicious flows with too-small-payload UDP packets
 (\autoref{tab:recall_by_anom}), without actually seeing packet payload, by 
 identifying malicious combinations of sMaxPktSz and sMinPktSz (maximum and minimum flow packet sizes, see
 \autoref{tab:data_flow_feat}).
The rationale is that we find malicious flows with too-small-payload UDP packets in our data 
 for SR2VP1 (other VIPs do not filter on payload sizes)
 are either made of all 56 or all 60-byte packets.
As a result, these malicious flows have only two possible combinations of
 sMaxPktSz and sMinPktSz (both 56 or both 60).
Since these two sMaxPktSz and sMinPktSz 
 combinations are rare among SR2VP1's UDP training flows 
 (0.01\% of 998M, not bad comparing to,
 for example, 0.46\% noises for benign destination ports),
 detecting flows with too-small-payload packets is equivalent
 to detecting flows with malicious sMaxPktSz and sMinPktSz
 combinations (both 56 and both 60).

\begin{figure*}
\centering
  \begin{subfigure}[b]{0.33\textwidth}
  \centering
  \includegraphics[width=0.99\textwidth]{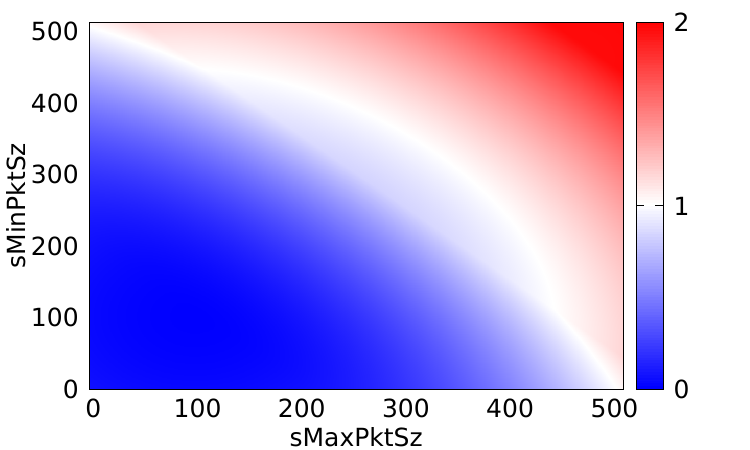}
    \caption{SR1VP1 (normal when both small)}\label{fig:range-pktsz-s1v1-loss}
\end{subfigure}
\begin{subfigure}[b]{0.33\textwidth}
  \centering
    \includegraphics[width=0.99\textwidth]{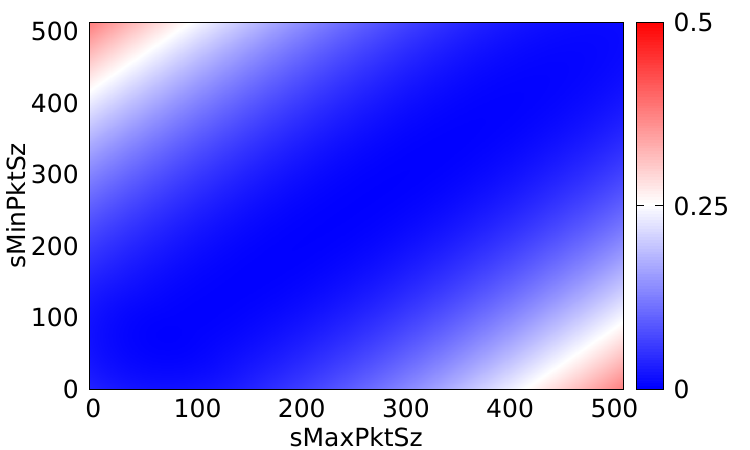}
  \caption{SR2VP1 (normal when similar)} \label{fig:range-pktsz-s2v1-loss}
\end{subfigure}
\begin{subfigure}[b]{0.33\textwidth}
  \centering
    \includegraphics[width=0.99\textwidth]{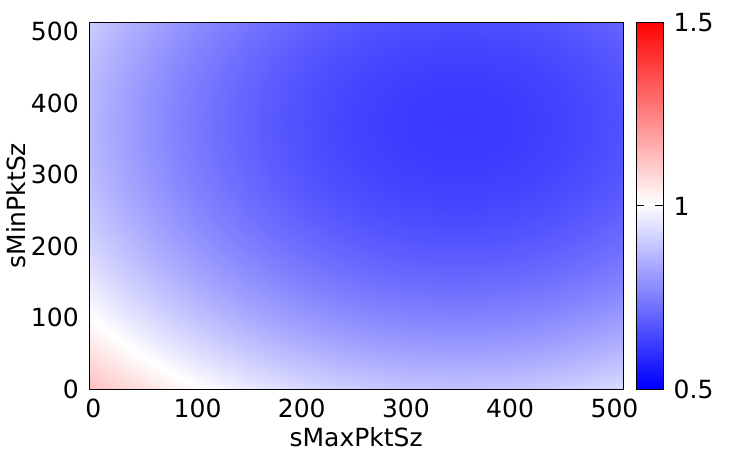}
  \caption{SR3VP1 (normal when both large)} \label{fig:range-pktsz-s3v1-loss}
\end{subfigure}
  \caption{Reconstruction Errors (Unit: $T_{det}$ from \autoref{eqn:det_thre}) for 1 Base Flow from 3 VIPs with Varying Packet Sizes (Byte) }\label{fig:range-pktsz}
\end{figure*}

\begin{figure*}
\centering
  \includegraphics[width=0.95\textwidth]{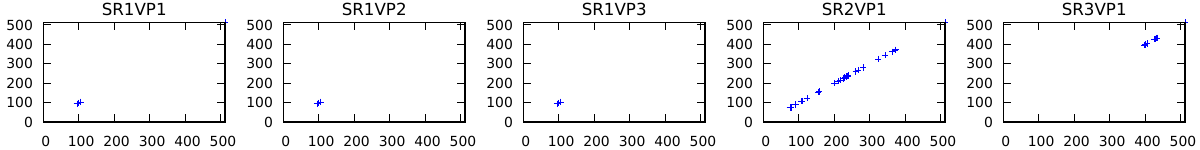}
    \caption{Frequent sMaxPktSz (X axis) and sMinPktSz (Y axis) Combinations in Training UDP Flows (At Least 1000 Occurrence).}\label{fig:train-minmax-pkt-occr}
\end{figure*}


We study the normality our models learn
 for sMaxPktSz and sMinPktSz combinations
 with counterfactual explanation.
We randomly draw 10 base flows from each of our 5 VIPs' test dataset,
 vary sMaxPktSz and sMinPktSz in base flows
 from 0 to 512 bytes (the largest packet size in
 our data) 
 with a step size of 1
 and watch how base flows' errors change.

We show our models learn incomplete normality for sMaxPktSz and sMinPktSz combinations, 
 by considering frequent combinations in training data,
 instead of all benign combinations (all except both 56 and both 60
 for SR2VP1 and all combinations for other 4 VIPs), as relatively normal.
Our models for SR1's three VIPs 
 consider base flows relatively normal when sMaxPktSz and sMinPktSz are both small
 (\autoref{fig:range-pktsz-s1v1-loss} shows reconstruction errors of one base flow from SR1VP1)
 because they mostly see small sMaxPktSz and sMinPktSz (at most 104 bytes)
 in training data (\autoref{fig:train-minmax-pkt-occr}).
The only exception is that we find a few training UDP flows for SR1VP1 
 (about 1.42\% of 999M) have large sMaxPktSz and sMinPktSz (both 512 bytes, 
 see blue pluses on top right corner of SR1VP1 chart in \autoref{fig:train-minmax-pkt-occr}).
However our model for SR1VP1 still considers sMaxPktSz and sMinPktSz of both 512
 abnormal (see the red on top right corner of \autoref{fig:range-pktsz-s1v1-loss}),
 likely treating these training flows 
 as noises.
%
%
Our model for SR2VP1 considers base flows relatively normal when sMaxPktSz and sMinPktSz
 are similar in value (\autoref{fig:range-pktsz-s2v1-loss} shows
 reconstruction errors for one  base flow from SR2VP1)
 because it mostly sees similar sMaxPktSz and sMinPktSz (both from 74 to 512 bytes) in training data (\autoref{fig:train-minmax-pkt-occr}).
Our model for SR3VP1 considers base flows relatively normal when sMaxPktSz and sMinPktSz
  are both large (\autoref{fig:range-pktsz-s3v1-loss} shows reconstruction errors for one base flows
 from SR3VP1)
 because it mostly see large sMaxPktSz and sMinPktSz (at least 397 bytes) in training data (\autoref{fig:train-minmax-pkt-occr}).

\subsection{Detecting Anomalies on Features}
	\label{sec:rlt_det_int_det_anom}
We next show that with correct normality, our
 models achieve reliable detection for anomalies on unordered feature
  (destination port).
With incomplete normalities, our models stills detect anomalies on unordered features
 (protocol and source port) with high recall but risk false positives. 
Lastly, we show that incomplete normalities could lead to low-recall detections for
 anomalies on ordered features (flow packet sizes).
%


\textbf{Unordered Destination Port:}
We show that the correct normality our models learn for destination port
 is key to the reliable detection of almost all flows with malicious destination ports
 (99.92\%; 1,260,943 out of 1,261,951, from \autoref{tab:recall_by_anom}).
Feature attribution analysis (\autoref{sec:md_feat_attr})
 confirms that most (98.55\%) of these true-positive detections
 are mainly triggered by anomalies from destination ports, providing in average $0.80\times$ threshold of 
 reconstruction errors.
For the rest reconstruction errors (in average $0.20\times$ threshold)
 needed, our models rely on anomalies on other features
 (mainly 
 Sport, sMaxPktSz, sMinPktSz and SIntPkt from \autoref{tab:data_flow_feat},
 with at least 10\% attributions in from 84\% to 3.53\%
 of these detections).

We argue that the tiny fraction of flows with malicious destination ports
 that our models miss (0.08\%, see \autoref{tab:recall_by_anom})
 are artifacts of our one-hot encoding of destination port
 rather than actual false negatives of our models' detection.
Recalling our models can not distinguish among adjacent 51 destination ports
 since we group them as one one-hot feature (\autoref{sec:md_data_ov}),
 our models consider these flows benign because their malicious destination ports are
 adjacent to the benign port (within 51).

\textbf{Unordered Protocol:}
Despite learning incomplete normality,
 we show our models still detect
 all 15,206 flows (\autoref{tab:recall_by_anom}) 
 with malicious protocols 
 (all except UDP, TCP and three other protocols, recalling \autoref{sec:rlt_det_int_learn_norm})
 by simply considering all flow with non-UDP protocol as 
 equally abnormal (see the horizontal blue line at normalized error of 1 for all non-UDP protocols in
 \autoref{fig:dist-reon-error-proto3v1}).
Feature attribution confirms that these true-positive detections
 are completely triggered by anomalies from protocols.

However by considering all non-UDP protocols equally abnormal,
 our models risk flagging flows with non-UDP benign protocols (such as TCP for
 SR1 and SR3) as malicious (false positives).
 causing the 28 false-positive detections our models made on test dataset (all TCP flows),
 recalling \autoref{sec:rlt_det_test_FP}.
%
%
%
%

\textbf{Unordered Source Port:}
Similarly, 
 our models detect most flows with malicious source ports (5,201 out of 5,334, 97.5\%, recalling \autoref{tab:recall_by_anom}),
 despite failing to learn the correct normalities,
 by considering all infrequent training source ports (including all but one malicious ports for SR1VP1
 and all malicious ports for other 4 VIPs) equally abnormal.
While our models risk false-positive detection
 by considering benign source ports that are infrequent in
 training data as relatively abnormal,
 we see no such false positives in test data 
 (as shown in \autoref{sec:rlt_det_test_FP}).

Feature attribution analysis confirms that
 most (99.79\%) of these true-positive detections 
 are mainly triggered by anomaly from source ports,
 providing in average about 0.79$\times$ threshold of reconstruction errors.
For the remaining reconstruction error needed (in average about 0.21$\times$ threshold),
 our models rely on anomaly from additional features (mainly
 sMaxPktSz, sMinPktSz, SIntPkt, SrcPkts, TcpOpt\_M
 and sTtl from \autoref{tab:data_flow_feat},
 with at least 10\% attribution
 in from 85.63\% to 1.6\% of
 these detections.)

We note that none of our models' 133 false-negative
 detection are due to they incorrectly consider one malicious source port from SR1VP1
 as relatively normal (recalling \autoref{sec:rlt_det_int_learn_norm}).
These false negatives are all due to our models
 cannot find enough anomalies from additional features
 (besides source ports) to trigger detection.

\textbf{Ordered Packet Sizes:}
%
We show that with incomplete normality for ordered features, our models risk low-recall detections.
While for unordered features like ports and protocols,
 our models consider values different from frequent training values
 as equally abnormal, 
our models consider values of ordered features 
 that are \emph{more different} 
 from frequent training values 
 as \emph{more abnormal} (see the gradual color changes from blue on left bottom to
 red on top right in \autoref{fig:range-pktsz-s1v1-loss}
 as an example).
As a result, our models
 risk considering malicious values for ordered features as relatively normal
 if they happen to be numerically close to 
 the frequent training values.

We show that with incomplete normality, our model for SR2VP1 incorrectly
 consider the malicious sMaxPktSz and sMinPktSz combinations (both
 56 and both 60 bytes) as relatively normal (shown as the blue bottom
 left corner of \autoref{fig:range-pktsz-s2v1-loss})
 because these malicious combinations happen to be numerically close to some
 frequent training combinations (both 74 bytes, as shown in SR2VP1's chart in \autoref{fig:train-minmax-pkt-occr}).
As a result, our models only detect a few flows with
 these malicious combinations (8.5\%, 215 out of 2,522, from \autoref{tab:recall_by_anom}).
Feature attribution analysis confirms that in these 215 true-positive detections,
 our model almost exclusively
 relies on anomalies from features
 other than sMaxPktSz and sMinPktSz.

\subsection{Using Anomalies from Multiple Features}
	\label{sec:rlt_det_int_muti_anom}

%
%

Lastly, we show that our models almost always combine anomalies from multiple flow features
 in detection, enabling our models to detect a quarter UDP flows with malicious packet payload contents
 (24.7\%, 2,036 out of 8,229, in \autoref{tab:recall_by_anom})
 even when they cannot see packet payload contents.
We breaking down number of features with significant attributions (at least 10\%)
 in our models' detection of all 1.2M malicious flows 
 in \autoref{tab:recall_by_anom}
 and show that our models uses multiple significantly-attributing 
 features in nearly all detections (99.90\% of 1.2M) 
 and uses 4 
 in most detections (79.16\% of 1.2M).
%
We argue that combining anomalies from multiple features is actually necessary for our models' detection
 by showing that almost all of these detected malicious flows would be missed (97.15\% of 1.2M) 
 if only using anomalies from the
 highest attributing features.

\section{Findings and Implications}
        \label{sec:imp}

We next distill our interpretation results 
  to three implications:
 training with noisy real-world data is possible (\autoref{sec:imp_noise}),
 autoencoder can reliably detect real-world anomalies on well-represented unordered features (\autoref{sec:imp_strength})
 and combinations of autoencoder-based AD and heuristic-based filters can help both (\autoref{sec:imp_combine}).

\subsection{Training with Noisy Data is Possible}
        \label{sec:imp_noise}

Our results show autoencoder-based AD models
 can be trained successfully on real-world data
 with noises,
 provided target features are \emph{well-represented}: 
 all benign values of target feature must appear frequently in
  the training data.
Our results support prior claim that attack-free training data
  does not exist outside simulation~\cite{Gates06a} 
  (we find some brief attacks in our training data),
  but we disprove the claim that noisy data makes AD training impossible.


We prove this ability to train on noisy data,
  showing that our
  autoencoder-based AD can learn normality in spite of noise.
For example, our models learn correct normality of destination port
 despite noise in the training data (0.46\% of 4.5M UDP flows have malicious destination port)
 because the benign port is the most frequent in training data (99.54\% of 4.5M
 UDP flows), recalling \autoref{sec:rlt_det_int_learn_norm}.

We also show that for under-represented features, 
 noise can be confused with normality,
 because both noise and some of their benign values
 are infrequent.
For example, 
 in \autoref{sec:rlt_det_int_learn_norm}, our models fail to learn
 benign protocol TCP (in 0.01\% of training flows) as normal likely due to our models
 consider TCP as noises, considering actual noises (in 0.22\% of training flows)
 are more frequent than TCP.
Our model for SR1VP1 learns
 a malicious source port (noise) as normal because this port 
 is the second most frequent 
 in training data (in 1.21\% of training UDP flows)
 and is 
 more frequent than almost all 
 benign sort ports.

\subsection{Autoencoder Reliably Detects Anomalies on Well-represented Unordered Features} 
       \label{sec:imp_strength}
Our results suggest that autoencoder-based AD could reliably detects real-world
 DDoS attacks, but only when all benign values for the DDoS flow's anomalous feature appear frequently in
 training data (so that our models could learn these benign values as normal)
 and when this anomalous feature is unordered (so that our models could crisply infer
 all the other values as abnormal).
If some benign values for this anomalous feature are infrequent in training data (as is usually the case
 for deny-listed features), our models risk considering these benign values as abnormal (false positives).
If this feature is ordered (such as packet rates, counts and sizes), even when all of its benign values appear frequently
 in training data, 
 our models still risk considering malicious values numerically close to 
 these benign values as normal (false negatives).


Our results thus refute the claim from prior work based on lab traffic
 that autoencoder-based AD detects DDoS attacks reliably (with true positive rate of 100\% and false positive rate of near 0)~\cite{N-BaIoT}.
Our results suggest that an autoencoder-based AD 
  will not detect sufficient attacks
  if it is the only DDoS-detection method
  in production environment.
%
We instead recommend using autoencoder-based AD as a compliment to 
 heuristic-based DDoS filters, see \autoref{sec:imp_combine}.

\subsection{Combine AD and Heuristic-Based Filters}
	\label{sec:imp_combine}

Finally, we show the potential for joint use of
  autoencoder-based AD and 
  heuristic-based filters (like in-house mitigation),
 since each has its own strengths.

We find our models are very good at finding and using anomalies from multiple features (4 in detection to
 most malicious flows \autoref{sec:rlt_det_int_muti_anom}).
ML-based AD is particularly important
  when
 the anomalies are not obvious to human perception,
  such as anomalies in 
   flow inter-packet arrival time, packet count and packet TTLs (recalling \autoref{sec:rlt_det_int_det_anom}).
However, our models are not very certain about each one of these anomalies
 and would miss
 97.15\% of its detected malicious flows if only using 
 the highest-attributing feature (\autoref{sec:rlt_det_int_muti_anom}).

The heuristic-base filter, by relying on human expertise, is very good at detecting
 malicious flow based on single anomaly.
(While in-house mitigation uses multiple heuristic-based filters, only one filter is used
 in each detection: the highest-priority filter triggered.)
For example, a flow with malicious destination port is certainly malicious because the server only 
 serves a short list of benign ports.
However we argue that it is more challenging for heuristic-based filters to make use of more subtle features to indicate malice,
 such as flow inter-packet arrival time or packet TTLs.
Our models are able to make use of these features (\autoref{sec:rlt_det_int_det_anom}),
  and can combine multiple suggestive features (\autoref{sec:rlt_det_int_muti_anom}).
%

We propose two possible strategies to combine 
 autoencoder-based AD and heuristic-based filters.
The first is to simply stack them: apply the heuristic filter first,
  to cover intuitive anomalies with great certainty.
Then add ML-based AD
  to covering additional anomalies
  that are not obvious or require combinations of features.
Our second strategy is to build new heuristics based on
  interpretations of what the autoencoder-based AD has discovered,
  as discussed in \autoref{sec:rlt_det_int}.
Such ``ML-discovered'' filters could directly use the ML model,
  or we could extract the relevant features into a new implementation.

\section{Related Work}
To the best of our knowledge, we are the first attempt to address 
 the two limitations (limited evaluation dataset and no detection interpretations) 
 in prior DDoS detection study using ML-based AD.

\subsection{DDoS Study using ML-based AD}
The most related class of prior work are those also detect DDoS attacks 
 with ML-AD models.

Most prior work in this class train some form of ML-AD models,
 such as one-class SVM models (\cite{ocsvm1,ocsvm2,ocsvm3}) and neural network 
 models 
 (\cite{N-BaIoT,DIOT,DIOT})
 with benign traffic and detect 
 attacks by looking for deviations from these 
 benign traffic.
Since these prior work mostly test their models with
 limited datasets including simulation~\cite{ocsvm3,Jun01a}, 
 lab traffic~\cite{ocsvm1,ocsvm2,N-BaIoT,DIOT} and DARPA/MIT dataset~\cite{ocsvm1},
 it is not clear how well their methods could work in real-world 
 production environment (\cite{Gates06a,Sommer10a}).
Moreover, they usually do not interpret their models' detection
 decision nor explore why their models work or not work in detecting 
 certain DDoS attacks.
In comparison, we evaluate our models with real-world benign and attack traffic 
 from a major cloud provider 
 and interpret 
 why our models work well on attacks of certain anomalies
 but not as well on the others.

K-means~\cite{kmean} and single-linkage~\cite{singlelinkage}
  have previously been used as clustering algorithms
to 
 separate benign and malicious traffic flows
 into different clusters.
Although their detection results are intuitively interpretable 
 (a flow is flagged as malicious since its features are qualitatively close to
 features of other flows in the ``malicious cluster''), 
they rely on manual inspection to determine which clusters
 are malicious. 
They also evaluate
 their methods with limited datasets (lab data~\cite{kmean} 
 and KDD datasets~\cite{singlelinkage}). 
In comparison, we do not rely on manual inspection for our detection, 
 and we test our methods on real-world traffic from a large cloud platform.

\subsection{DDoS Study using Other Techniques}
Many prior work detect DDoS attacks with other techniques. 
We classify them into following 3 classes.

\textbf{ML-based binary classification:}
This class of papers train some form of ML binary classification models
 (such as KNN~\cite{nn6}, decision tree~\cite{decesion_tree1,nn6},
 two-class SVM~\cite{tcsvm1,tcsvm2}, 
 random forest~\cite{nn6} and
 neural network models~\cite{nn1,nn2,nn3}) 
 with both benign and attack traffic.
These ML models thus identify attacks similar to the ones they 
 have seen during training. 
In comparison, we focus on a different model (ML-AD model)
 and by training with only benign traffic and looking for deviations
 from these benign traffic, our models
 do not rely on on knowledge of known attacks
 and have the ability to identify potential unknown attacks.

\textbf{Statistical AD:}
This class of papers apply statistical models
 (such as adaptive threshold~\cite{adathre1}, cumulative 
 sum~\cite{cusum1,adathre1}, 
 entropy-based analysis~\cite{entropy1} 
 and Bayesian theorem~\cite{Bayesian1}) to 
 identify abnormal traffic pattern that is significantly different 
 from some or all of previously seen (benign) traffic pattern.
These papers thus could also cover potentially unknown attacks. 
In comparison, we focus on AD based on 
 ML models instead of statistical models.

\textbf{Heuristic-based rule:}
This class of papers use heuristic-based rules to detect 
 specific types of attacks matching their heuristics. 
For example, history-based IP filtering remembers frequent remote IPs 
 during peace time and consider traffic from other IPs during attack time as
 potential DDoS traffic~\cite{hurestic1}. 
Hop-count based filtering identifies spoofed DDoS packets 
 by remembering peace-time IP to (estimated) hop count mapping and 
 considering packets with unusual IP-to-hop-count mapping 
 during attack time as spoofed DDoS packets~\cite{hurestic2}.
 In comparison, we use a different method
(ML-based AD) and could cover many different types of attacks
instead of only a specific type.

\section{Conclusion}

This paper addresses two limitations in prior studies
  of ML-based AD:
  use of real-world data, and interpretation of why the models are successful.
We apply autoencoder-based AD to 57 real-world DDoS 
 events captured at 5 VIPs of a large commercial cloud provider.
We use feature attribution and counterfactual techniques to explain when
  our models work well and when they do not.
Key results are that our models detect most, if not all,
 malicious flows to 4 VIPs under attacks, with near-zero false positives.
Interpretation shows 
 our models are better at detecting anomalies on allow-listed features than 
 those on deny-listed features
  because our models are more likely to learn correct normality for allow-listed features.
We then show that our models are better at detecting anomalies on unordered
 features than those on ordered features because even with incomplete normality,
 our models could still detect anomalies on unordered feature with high
 recall.
Key implications of our work are that
  training with noisy data is possible,
  that autoencoder-based AD can reliably detect anomalies on well-represented unordered
  features
  and that autoencoder-based AD and heuristic-based filters have complementary strengths.

\section*{Acknowledgments}

We thank Yaguang Li from Google, Wenjing Wang from Microsoft 
 and Carter Bullard from QoSient for their comments on this paper.

This work was begun with the support of a summer internship by Microsoft.

Hang Guo and John Heidemann's work in this paper
  is based on research sponsored by Air Force Research Laboratory under
agreement number FA8750-17-2-0280. 
The U.S.~Government is authorized to reproduce and distribute
reprints for Governmental purposes notwithstanding any copyright
notation thereon.

\bibliographystyle{abbrv}
\begin{spacing}{0.9}
\small{
\bibliography{paper.bib}
}
\end{spacing}

\end{document}